\documentclass[ijgi,article,submit,moreauthors,pdftex,10pt,a4paper]{mdpi} 
\usepackage{color, colortbl}

\firstpage{1} 
\articlenumber{x}
\doinum{10.3390/------}
\pubvolume{xx}
\pubyear{2018}
\copyrightyear{2018}
\externaleditor{Academic Editor: name}
\history{Received: date; Accepted: date; Published: date}
\pdfoutput=1

\Title{Mobile phone indicators and their relation to the socioeconomic organisation of cities.}

\address[a]{CNRS, Centre Maurice Halbwachs, UMR 8097, Paris, France.}
\address[b]{University College London, Centre for Advanced Spatial Analysis, London, UK.}
\address[c]{University of Newcastle, Open Lab, Newcastle upon Tyne, UK.}
\address[d]{Orange Labs France, SENSe, Châtillon, France.}

\Author{Clémentine Cottineau $^{1,2,\ddagger,}$* and Maarten Vanhoof $^{3,4,\ddagger}$}
\AuthorNames{Clémentine Cottineau and Maarten Vanhoof}

\address{%
$^{1}$ \quad CNRS, Centre Maurice Halbwachs, UMR 8097, Paris, France; clementine.cottineau@ens.fr\\
$^{2}$ \quad University College London, Centre for Advanced Spatial Analysis, London, UK\\
$^{3}$ \quad University of Newcastle, Open Lab, Newcastle upon Tyne, UK; m.vanhoof1@ncl.ac.uk\\
$^{4}$ \quad Orange Labs France, SENSe, Châtillon, France}

\corres{Correspondence: clementine.cottineau@ens.fr; Tel.: +33 (0)1 80 52 14 42}

\secondnote{These authors contributed equally to this work.}

\abstract{Thanks to the use of geolocated big data in computational social science research, the spatial and temporal heterogeneity of human activities are increasingly being revealed. Paired with smaller and more traditional data, this opens new ways of understanding how people act and move, and how these movements crystallise into the structural patterns observed by censuses. In this article we explore the convergence of mobile phone data with more classical socioeconomic data from census in French cities. We extract mobile phone indicators from six months worth of Call Detail Records (CDR) data, while census and administrative data are used to characterize the socioeconomic organisation of French cities. We address various definitions of cities and investigate how they impact the relation between mobile phone indicators, such as the number of calls or the entropy of visited cell towers, and measures of economic organisation based on census data, such as the level of deprivation, inequality and segregation. Our findings show that some mobile phone indicators relate significantly with different socioeconomic organisation of cities. However, we show that found relations are sensitive to the way cities are defined and delineated. In several cases, differing city definitions delineations can change the significance or even the signs of found correlations. In general, cities delineated in a restricted way (central cores only) exhibit traces of human activity which are less related to their socioeconomic organisation than cities delineated as metropolitan areas and dispersed urban regions.}


\begin{document}

\section{Introduction}

\subsection{The single-city focus of urban sensing}

The quantitative analysis of mobile phone records\citep{gonzalez2008understanding}, smart card traces \citep{zhong2016variability}, or credit card transactions \citep{lenormand2015influence, de2015unique}, is increasingly revealing the regularities behind human daily practices, such as mobility or social interactions (e.g. \citep{alessandretti2018evidence, pappalardo2015returners}), very often in an urban context. The main advantages of sensed big data are well known and consist of, among others, the reduction of collection and treatment cost, the increase of sample sizes, and the possibilities for more timely and recurring observations. In the case of mobility studies, for example, \citet{batran2018inferencing} note that: "While traditional survey methods provide a snapshot of the traffic situation in a typical weekday, mobile phone data can capture weekday and weekend travel patterns, as well as seasonal variation of a large sample of the population at a low cost and wide geographical scale". In contrast, the disadvantages of sensed big datasets are that they suffer from spatial and temporal sparseness \citep{lu2017understanding,Vanhoof2018HomeDet}, from lack of - or an unknown degree of - representativeness \citep{longley2015geotemporal, arai2016comparative}, and from issues regarding anonymity \citep{de2015unique}.

Within the urban sensing literature, mobile phone data play a prominent role as they form a source of passively collected information (users do not need to make an explicit action to share their locations as would be the case in, for example, location-based services or social networks), for large shares of populations (high shares of the world population now owns a mobile device of any sort), captured at a rather high spatial resolution (in general, the density of cell towers is high in urban areas). Mobile phone data research in an urban context has been applied to a diversity of individual cities, or to international comparison of cities: Paris \citep{schneider2013unravelling}; Maputo \citep{batran2018inferencing}, Dhaka \citep{lu2017understanding}, Santiago \citep{dannamann2018time}, Boston and Singapore \citep{xu2018human}, London, Singapore and Beijing \citep{zhong2016variability}. Research with a focus on a single city, or a set of single cities, bears the advantage that it can easily tap into local knowledge when questioning the obtained, quantitative results. This leads to better insights that can be used in urban planning and policy.

\subsection{Mobile phone indicators}

One problem of the single-city focus in previous research is that it can not ensure that observations made in one city (usually a capital city of large population) remain valid for other cities. As a consequence, it is unclear whether findings can be generalized over different types of cities. The creation of mobile phone indicators avoids this problem. Since mobile phone indicators are typically calculated for large user samples covering multiple cities, aggregation of individual-level indicators in space allows to compare findings between cities. In addition, mobile phone indicators can be paired with other datasets so that multi-variate methods can support interpretation. In the case of mobile phone data, creating individual indicators is possible at a nation-wide scale (as datasets are mostly provided by national operators) but it is not a straightforward task. For example, differences in the spatial resolution of observations make it hard to create comparable indicators for individual mobility \citep{Vanhoof2018MobEntr}, and it is known that home detection methods, which enable the spatial allocation and aggregation of individual users, still face severe challenges when it comes to validation and error estimation \citep{Vanhoof2018HomeDet,Vanhoof2018Sensitivities}. 

Regardless some methodological challenges, creating mobile phone indicators and pairing them with census data is deemed promiscuous by multiple official statistics offices and has been performed in academic literature on several occasions. \cite{Pappalardo2016}, for example, show how in France a mobile phone indicator on the diversity of movement (the mobility entropy) relates directly to the European Deprivation Index (EDI), \cite{Eagle2010} describe the relation between regional calling patterns and economic development in the UK, \cite{Decuyper2014} discuss the relation between calling and purchase behavior and food security in a Central African country, and \cite{Frias-martinez2013} investigate relations between several mobile phone indicators (call, movement and purchase behavior) and multiple census variables on education, demographics and purchase power in a Latin American country. With the exception of \cite{Vanhoof2018MobEntr}, who study relations between mobile phone and census indicators for different urban areas in France, one clear shortcoming of these studies is that their analyses are fixed on the nation-level only, leaving a missed opportunity to explore the empirical relations between human mobility, social interactions, and the socioeconomic organisation of cities. 

\subsection{Sensitivity of urban scaling laws to city definitions}

When intending to compare values of mobile phone indicators between cities, it is important to have a clear definition of what is considered a city. This is especially true since recent works on urban scaling and Zipf's law of census data \citep{Arcaute2015,veneri2016city, cottineau2017diverse} have shown that the delineation of cities can substantially influence results and interpretations, mainly because areas either included or not in different delineations have heterogeneous properties. Despite the fact that this issue is traditionally overlooked (sometimes for good reason, because data is only available for a single delineation), it is to suspect that average human activity sensed in general, and mobile phone indicators in particular, are similarly sensitive to city delineation.

Proceeding one step further to the relations between mobile phone indicators and census data, one can ask themselves the question to which degree such relations will be influenced by city definitions. Indeed, what is unclear from previous work on mobile phone indicators is how statistical relations to census indicators, whether obtained from multi-variate analysis at nation level or in the form of urban scaling laws, are determined by the way cities are defined. Before this question gets answered, empirically produced relations will be insufficiently trustworthy to properly engage with theoretical hypotheses such as the ones about the link between mobility, human interactions, and the socioeconomic organisation of cities.

\subsection{Research question and relevance}
Consequently, in this paper, we explore to which degree relations between mobile phone indicators and census indicators are sensitive to the chosen city definition, this is, to the way we delineate cities. Doing so, we question the value of mobile phone indicators on mobility and human interactions for the understanding of the socioeconomic organisation of cities. To test sensitivity to city definitions, we run a parametric simulation of different city delineations in France. Assuming that mobile phone indicators depict different types of spatial variation compared to socioeconomic urban indicators (e.g. calling patterns might be less influenced by infrastructural elements and built-up environment and therefore more homogeneously spread across the country than, for instance, wages) and building upon recent empirical work that highlights the influence of city definition when assessing scaling laws \cite{Arcaute2015,cottineau2017diverse}, our hypothesis is that city definitions will influence empirical relations between the two types of indicators in a non-trivial way. As discussed before, multiple works have uncovered relations between mobile phone indicators and census data but, to the best of our knowledge, all of them do so based on one city definition only. If relations are sensitive to city definitions, this would have considerable implications for their validity, interpretation, and potential use in (predictive) applications.

\section{Data}

This section introduces the census and mobile phone indicators we will use in our investigation. We limit the analysis of census data to three socioeconomic urban indicators chosen for their social relevance and easy interpretation. They relate to three dimensions of the economic organisation of cities: their level of poverty (or deprivation), their level of inequality (distribution of wages) and their level of segregation (spatial distribution of wages) and are introduced in \ref{subsec:census_data}. Regarding mobile phone indicators we deploy 15 mobile phone indicators derived from a mobile phone dataset in France and covering aspect of human mobility and social interaction. Their creation and properties are described in \ref{subsec:phone_data}.

\subsection{Census data on segregation, inequality, and deprivation}\label{subsec:census_data}
We limit the analysis of census data to three socioeconomic urban indicators chosen for their social relevance and easy interpretation. They relate to three dimensions of the economic organisation of cities: their level of poverty (or deprivation), their level of inequality (distribution of wages) and their level of segregation (spatial distribution of wages).

The deprivation indicator is measured by the European Deprivation Index (EDI) created for France by \cite{Pornet2012}. The EDI is an individual deprivation indicator constructed from an European survey specifically designed to study deprivation \cite[p.~982]{Pornet2012}. It is created as a composite measure incorporating information on both subjective and objective poverty and the attribution of the weights for different contributing factors is done specifically for France. 

Wage inequality at the level of the city was computed similarly to \cite{Cottineau2018Defining}. The inequality index is calculated using the Gini index method \cite{fuller1979estimation} on groups of similar wage earners as described in the CLAP database holding information on French firms and establishments. The inequality index measures the overall dispersion in the distribution of wages at the city level, and varies between 0 (extreme equality) and 1 (extreme concentration of wages).\\
Wage segregation for cities was also computed similar to \cite{Cottineau2018Defining}, using the \cite{reardon2009measures}'s $R^O$ index of ordinal segregation for classes of wages retrieved, again, from the CLAP database. This segregation indicator measures the spatial dispersion of the distribution of wages between communes of the city, and varies between 0 (homogeneous city) and 1 (extreme segregation by wages in the city).

\begin{figure}[h!]
\medskip
\centering
\resizebox{0.3\textwidth}{!}{ 
\includegraphics{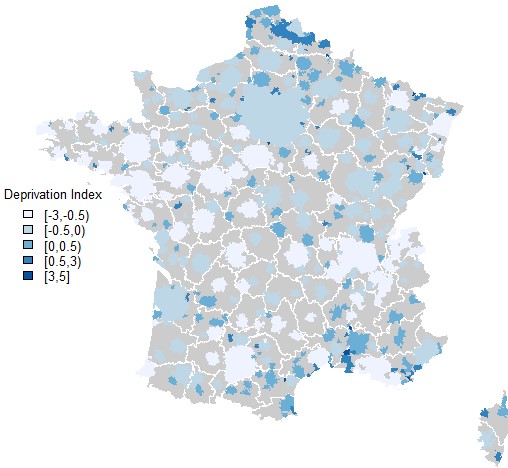}
}
\resizebox{0.3\textwidth}{!}{ 
\includegraphics{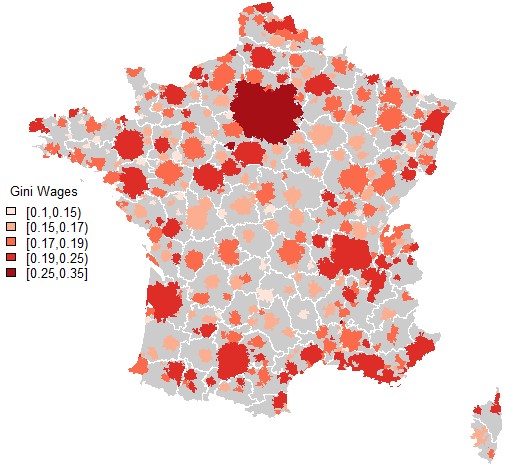}
}
\resizebox{0.3\textwidth}{!}{ 
\includegraphics{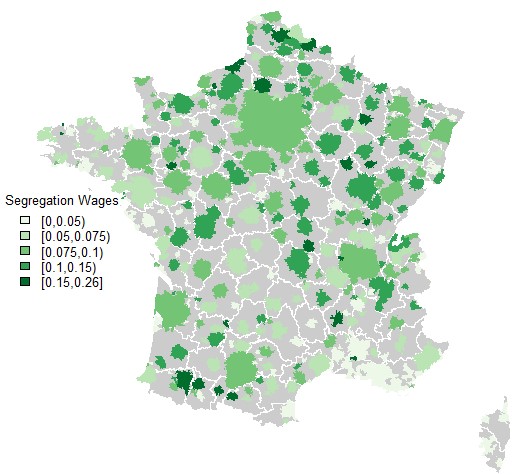}
}
\caption{Maps of deprivation, inequality and segregation levels for metropolitan areas with more than 10,000 residents in 2011.}
\label{fig:au_maps_socioeco}
\end{figure}

The distribution of deprivation, inequality and segregation in French metropolitan areas (Aires Urbaines) is depicted on figure \ref{fig:au_maps_socioeco}. It shows different spatial logics (size effects for inequality, which is higher in large cities and a regional differentiations for deprivation levels and segregation, which are higher in Northern cities for example).

\subsection{Constructing indicators from the French mobile phone data}\label{subsec:phone_data}
To create mobile phone indicators, we use a French mobile phone dataset collected during the period between May $13^{th}$ and October $15^{th}$ 2007. The dataset is owned by Orange and holds information from the Orange cellular network, which in 2007 consisted of about 18,275 cell towers nation-wide. The dataset itself consists of \textit{Call Detailed Records (CDR)}, that collect information on the time, deployed cell tower, initializing user, receiving user and duration/length of each call or text made by about 18 millions Orange subscribers. The French CDR dataset has been extensively studied before \cite{Grauwin2017,Pappalardo2016,Vanhoof2017domestictourism,Vanhoof2017miningurbanareas,Vanhoof2018HomeDet,Vanhoof2018MobEntr,Vanhoof2018Sensitivities} and is one of the largest datasets that guarantee access to individual user data over such a long time period.

To construct mobile phone indicators from CDR data, a version of the open-source python library Bandicoot \cite{DeMontjoye2016} was implemented on the big data infrastructure of the Orange Labs France. For each user and for each month in the observation period, a set of indicators (table \ref{table:indicators}) is calculated. Because indicators at the individual level entail a small but potential privacy risk, user indicators are aggregated at cell tower level. Aggregation is done for each user and for each month based on the assumed home location according to a home detection algorithm. We tested two home detection algorithms: the maximum amount of activities and the distinct days algorithm as defined by \cite{Vanhoof2018HomeDet,Vanhoof2018Sensitivities}. The result, for each home detection method, is a distribution of values for all indicators for each cell tower in the Orange cell network. When comparing results of using the distinct days algorithm (home is the cell tower where the user was present the maximum number of distinct days during a month) to results when using the maximum activities algorithm (home is the cell tower where the user made most mobile phone actions during a month), no substantial differences are found. This is not entirely surprising as results in \cite{Vanhoof2018HomeDet,Vanhoof2018Sensitivities} already suggested performance of both algorithms to be only slightly different when comparing population counts at nation-level, with the performance of both algorithms being best in September \cite{Vanhoof2018HomeDet}. Therefore, we choose to present the remaining results using the distinct days algorithms, limiting our analysis to September 2007 only. 

\begin{table}[ht]
\caption{Description of mobile phone indicators}
\centering 
\begin{tabular}{|p{5.5cm} | p{9cm}|}
\hline 
\textbf{Mobile phone indicator} & \textbf{Description }\\[0.5ex] 
\hline 
\rowcolor[gray]{.9}Number of calls & Number of calls made or received \\
Active days & Number of active distinct days  \\
\rowcolor[gray]{.9}Percentage nocturnal calls & Percentage of calls made between 7pm and 9am \\
Duration of calls & Mean, median, or standard deviation of duration of all calls \\
\rowcolor[gray]{.9}Inter-event time &  Mean, median or standard deviation of the duration between consecutive calls \\
Number of contacts & Number of contacts interacted with \\
\rowcolor[gray]{.9}Interaction per contact & Mean, median or standard deviation of the amount of interactions per contact \\
Entropy of contacts* (eq.\ref{eq:ent_contacts})& Entropy measure of calls to contacts \\
\rowcolor[gray]{.9}Number of visited cell towers & Number of cell towers used to make calls \\
Radius of gyration* (eq.\ref{eq:rad_gyr}) & Radius of gyration of movement patterns based on visited cell towers \\
\rowcolor[gray]{.9}Entropy of visited cell towers* (eq.\ref{eq:ent_towers}) & Entropy measure of visited cell towers \\
Distance between l1 and l2 & Distance between most plausible and second most plausible 'home' cell tower given a home detection algorithm \\
\rowcolor[gray]{.9}Spatial uncertainty & Uncertainty measure of the detection of the most plausible home location \\
Calls at home & Number of calls made at the presumed home cell tower  \\
\rowcolor[gray]{.9}Percentage calls at home & Percentage of calls made at the presumed home cell tower  \\
[1ex]
\hline
\end{tabular}
\label{table:indicators} 
\end{table}

The definition of most of the mobile phone indicators is straightforward, but some merit a proper mathematical definition (* in table \ref{table:indicators}). The radius of gyration for example, is a measure of a user's mobility surface defined as the spatial spread of the cell towers visited by a user relative to his or her centre of mass, which is defined as the mean point of all his/her visited cell towers \cite{Pappalardo2016}:
\begin{equation}\label{eq:rad_gyr}
Radius\:of\:gyration= \sqrt{{\frac{1}{N}} \sum_{i\in L} n_i (r_i - r_{cm})^2}
\end{equation}
where $L$ is the set of cell towers visited by the user, $n_i$ is each cell tower's visitation frequency, $N= \sum_{i\in L} n_i$ is the sum of all the single frequencies, $r_i$ and $r_{cm}$ are the vector coordinates of cell tower $i$ and centre of mass respectively. 

The entropy of visited cell towers is a reflection of the diversity of user's movement pattern. It is defined as the Shannon entropy of the pattern of visited cell towers \cite{Pappalardo2016,Vanhoof2018MobEntr}:
\begin{equation}\label{eq:ent_towers}
Entropy\:of\:cell\:towers=-{\frac{\sum_{l \in L} p(l) \log p}{\log N}}
\end{equation}

where $L$ is the set of cell towers visited by the user, $l$ represents a single cell tower, $p(l)$ is the probability of a user being active at a cell tower $l$, and $N$ is the total number of activities of one user.

The calculation of the entropy of called contacts is identical to the entropy of visited cell towers, only here, the pattern of called contacts replaces the visited cell towers \cite{Pappalardo2016}:
\begin{equation}\label{eq:ent_contacts}
Entropy\:of\:contacts=-{\frac{ \sum_{e \in E} p(e) \log p}{\log N}}
\end{equation}
where $C$ is the set of all contacts of a user, $c$ represents a single contact, $p(c)$ is the probability that the user is contacting a contact $c$ when active, and $N$ is the total number of activities of one user.

\begin{figure}[h!]
\medskip
\centering
\resizebox{1\textwidth}{!}{
\includegraphics{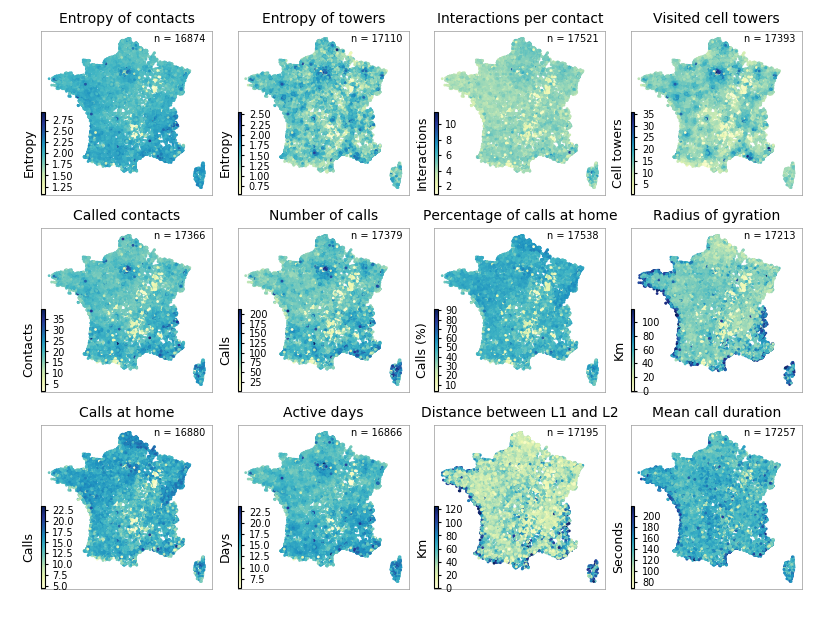}
}
\caption{Maps of several mobile phone indicators at cell tower level. Indicators are calculated for the month September 2007, for all users in the French CDR dataset. Users are aggregated at cell tower level by the Distinct Days Algorithm. Each dot on the map is a cell tower and displays the average indicators value of all users designated to this cell tower. Cell towers with an average value that is higher, or lower, than 3 standard deviations from the nationwide average are omitted, hence the differing number  of displayed cell towers ($n$) between maps.}
\label{fig:maps_mob_ind}
\end{figure}

Spatial patterns of the calculated mobile phone indicators for September 2007 (aggregated by average per cell tower) are presented in figure \ref{fig:maps_mob_ind}. Most spatial patterns show a clear urban-rural dichotomy with, for example, number of active days, called contacts, visited cell tower and number of calls being higher in city centres where, most likely, mobile phone use was more adopted by the general population compared to rural areas\footnote{An other explanation for the number of antennas visited might be that their distribution is more dense in cities, thus artificially raising the value for a similar surface travelled by users. However, this bias was taken into account for the entropy of visited cell towers.}. The spatial pattern of the radius of gyration and distance between L1 and L2 stand out but are not unsurprising. Here, cell tower averages are influenced by users performing domestic tourism (see also \cite{Vanhoof2017domestictourism}) that have a detected home at the seaside (L1) and a second plausible home location(L2) further away, resulting in a high radius of gyration value and a high L1-L2 distance. Another intriguing pattern is visible in the absolute number of calls at home. This pattern remains unexplained but could point to differences in the adoption of mobile phone technology between regions. It is interesting, however, to note that this regional pattern dissolves when looking at the percentage of call at homes. The relative number of calls at homes is rather uniformly distributed in France except for the extreme remote and rural areas where almost 100 \%  of calls are performed at home.

Similar to the socioeconomic indicators, the distributions of mobile phone indicators exhibit different spatial patterns \ref{fig:maps_mob_ind}, from a metropolitan concentration (active days, number of calls, mobility entropy) to a coastal concentration (radius of gyration, mean call duration).

\section{Methods}

In this section we'll explain the methods used to investigate the sensitivity to city definitions of the relations between mobile phone indicators and census data. \ref{subsec:aggregation} explains how we align both type of indicators as they are gathered at a different spatial resolutions. The following sections explain how we simulate 4914 different city definitions \ref{subsec:simulating_cities}, how we calculate correlations between indicators for each of these city definitions \ref{subsec:correlations} and how we represent the obtained results \ref{subsec:representing_results}. In a final section we propose a way to capture the 4914 different city delineations into a more interpretable set of 6 classes of city definitions. And we show how, for each of these classes, there exists a limited relation between the three census indicators we use to describe the socioeconomic situation of cities, pointing out their independence against one another.

\subsection{Aggregation: from cell towers to commune to city}\label{subsec:aggregation}
    In order to compare mobile phone indicators with census data, we need to find a common perimeter to aggregate the data. Since the majority of census data is available within boundaries defined by administrative units ({\it communes} in France), we choose to extrapolate the mobile phone indicators (available at the cell tower level) to match the {\it communes} boundaries. There is no information about the exact perimeter each cell tower covers but it is reasonable to assume that phones will log in to the closest antenna available at all time, thus drawing coverage areas close to Voronoi polygons around the cell towers. 

After building the layer of Voronoi polygons, we intersect it with the layer of {\it commune} polygons using the programming language R. In each of the resulting intersection polygons, we computed the share of area that the intersection polygon represents with respect to its original cell tower Voronoi polygon. We allocated the number of users from the original cell tower Voronoi polygons to the intersections based on the share of area they represented. Finally, the data for {\it communes} were aggregated based on the number of users in each intersection polygon and a weighted average of all the indicators of mobile phone activity based on the share of users, with respect to the {\it commune} total users.

\subsection{Simulating city definitions}\label{subsec:simulating_cities}
Now that we have mobile phone indicators and census data available at the {\it commune} level, we can simulate different city definitions by grouping {\it communes} together, and aggregating their indicators. 
We use a generative, parametric method to simulate a range of city definitions. This method is inspired by the official city definitions in France (which define a city centre based on a minimum density, a periphery based on a minimum share of commune dwellers commuting to the city centre, and then apply a population minimum) and has been produced by \cite{cottineau2017diverse}. The method simulates different city definitions by aggregating the French communes into a set of cities by iterating over three parameters: a density minimum $d$ to define city centres (from 1 to 20 persons per ha, by steps of 0.5), a minimum percentage $f$ of workers in a {\it commune} commuting to this city centre (from 0 to 100\%, by steps of 5) and a minimum population $p$ within the resulting city (from 0 to 50,000, by steps of 10,000). In total, the simulation renders 4914 different city definitions, i.e. spatial delineations of aggregated {\it communes}  (4914 = 39 density thresholds x 21 flow thresholds x 6 population thresholds). For each city definition we compute for all cities, the total population considered, the overall inequality (Gini coefficient of wage groups present in the city), the spatial segregation (of wages groups in the {\it communes}, the average deprivation index (EDI) and the weighted average of the mobile phone indicators based on the {\it communes} values with respect to their number of users.

\subsection{Correlations between mobile phone indicators and census data}\label{subsec:correlations}

Having prepared mobile phone and census indicators, as well as a method to simulate different city definitions, we investigate the correlations between mobile phone and census indicators, and their sensitivity to city definition. Specifically, we investigate the relation between the three census variables (Gini index, Segregation index and EDI) and all mobile phone indicators in table \ref{table:indicators} for each of the 4914 city definitions. For each combination of census indicator, mobile phone indicator, and city definition, the Spearman correlation coefficient is calculated based on the point cloud of all cities adhering to the deployed city definition. The Spearman correlation is preferred over the Pearson correlation as the latter is mainly for linear relationships which is not verified for in the automated computation we performed. In section \ref{subsubsec:multiple_regression}, we combine the effect of all three aspects of the socioeconomic organisation of cities into a multiple linear regression of each mobile phone indicator. 

\subsection{Representing results for all 4914 city definitions}\label{subsec:representing_results}

Representing the resulting correlations for 4914 city definitions will be done by value distributions over all 4914 city definitions (\ref{subsubsec:distr}) or by heatmaps when discussing the impact of the simulation parameters $d$, $f$, and $p$ (\ref{subsubsec:heatmaps}). The coordinates of the heatmaps are made up of the different thresholds for population ($p$), density ($d$) and flow ($f$), or thus the different parameters of the city definition, and are coloured according to the obtained Spearman correlation coefficient (or related r$^2$). In this way they offer a more expressive view of the correlation and their sensitivity to different city definitions. 

When discussing in depth the relation between the different socioeconomic variables and mobile phone indicators (\ref{subsubsec:multiple_regression}), we will reduce the number of studied city definitions to a manageable 6 case studies. These 6 case studies correspond to the centres of classes formed by thousands of city definitions. The clustering of city definitions was performed on a fixed population minimum ($p$) of 10,000 residents (the threshold  most frequently used in urban system studies \cite{cottineau2017metazipf}), because we want its focus to be on the density ($d$) and flow ($f$) thresholds that produce a variation in the spatial extent of cities amongst city definitions). Fixing the population minimum ($p$) on 10,000 residents, the clustering is thus performed on 819 definitions only. 

\subsection{Clustering city definitions to 6 classes}
\label{subsec:clustering}

\begin{figure}[h!]
\centering
\caption{6 classes of city Definitions (example with Population > 10,000}
\label{fig:zonesDefs}
\resizebox{0.8\textwidth}{!}{ 
\includegraphics{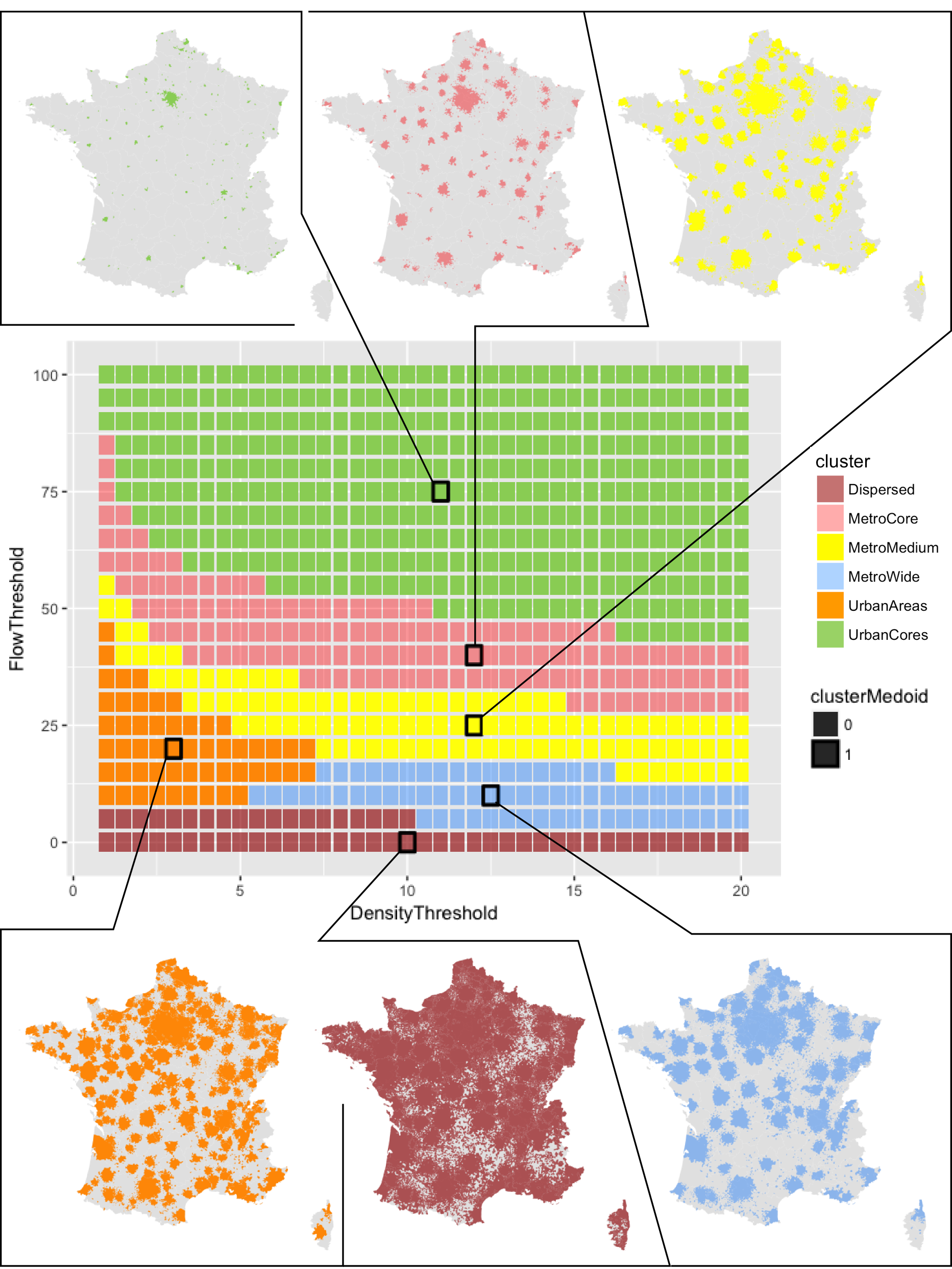}
}
\end{figure}

Clustering is based on the similarity of commune membership in cities over different city definitions. Starting from the membership table of communes to cities in the different definitions, we compute a dissimilarity matrix of city definitions based on their vector of about 36000 asymmetric binary values (indicating if each commune is included or not in a city) and a Gower distance \citep{gower1971general}. We then apply a k-medoid clustering \citep{kaufman1987clustering} algorithm to the dissimilarity matrix and, judging from the silhouette width and the groupings obtained, we identify 6 classes (figure \ref{fig:zonesDefs}) and their centroid. 

The most numerous definitions ("Urban cores" in green on figure \ref{fig:zonesDefs}) represent ways of defining cities which results in very small aggregates. These "urban cores" are obtained either by highly dense communes with little periphery (right of figure \ref{fig:zonesDefs}, i.e. high density threshold and middle flow threshold) or a wider centre with no periphery (top left, i.e. middle density threshold and high flow threshold), similarly to {\it Unités Urbaines} as defined by the French statistical office INSEE. This is the most restrictive way of thinking about cities and the centroid for this class is obtained with a density minimum of 11 persons per ha and a minimum of 75\% of commuters from peripheral communes working in the centre.

The next three classes (going clockwise on figure \ref{fig:zonesDefs}) stretch through similar values of density minima but have increasing peripheries (by lowering the percentage needed to attached communes to the metropolitan centres). We call them "MetroCores", "MetroMedium" and "MetroWide". Their centroids have a similar density minimum (12-12.5) and a decreasing flow minimum (40\%, 25\% and 10\%).

With an even lower flow minimum on average, the "Dispersed" class is the furthest away from a strict view on cities. Indeed, almost all French communes are included in a "city" of some sort according to this definition, as only a few commuters are sufficient to aggregate peripheral communes to high population centres. This class is a limit case, represented by a centroid of density minimum of 10 and a flow minimum of 0.

Finally, the "Metropolitan Area" class includes definitions closest to that of the { \it Aires Urbaines} as defined by the French statistical office INSEE. They are characterized by relatively low density thresholds (from 1 to 8.5 residents per ha) and a limited range of flow thresholds (from 10 to 45\%). The city definitions in this class, as exemplified by the centroid case, are generally more numerous than in the "Metro-*" cases, have always some periphery but exclude the most rural parts of the country.

\subsubsection{Correlation between socioeconomic indicators for centres of definition classes}\label{subsubsec:distrMedoids}

Before continuing on the correlation between mobile phone indicators and indicators of the socioeconomic organisation of cities, we want to check that the three indicators we picked to represent this organisation are independent from one another. 

\begin{table}[ht]
\caption{Spearman correlations between three census indicators for the centroid definition of all the six classes.}
\centering 
\begin{tabular}{|l|c|c|c|}
\hline 
\textbf{} & \textbf{Deprivation - Inequality } & \textbf{Inequality - Segregation} & \textbf{Segregation - Deprivation}\\[0.5ex] 
\hline 
\rowcolor[gray]{.9} \cellcolor[HTML]{F8A102}{\color[HTML]{333333} UrbanAreas} & 0.060       & -0.082         & -0.028  \\
\cellcolor[HTML]{C38A8A}Dispersed                         & 0.062    & \textbf{-0.252} & -0.144   \\
\rowcolor[gray]{.9}\cellcolor[HTML]{AEDD4C}UrbanCores                        & -0.044      & 0.192           & 0.186 \\
\cellcolor[HTML]{FFFE65}MetroMedium                       & -0.059       & -0.047         & -0.156 \\
\rowcolor[gray]{.9}\cellcolor[HTML]{FFCCC9}MetroCore                         & -0.072        & 0.119       & -0.131  \\
\cellcolor[HTML]{B0DCF0}MetroWide                         & -0.041      & -0.156     & -0.069 \\
[0.1ex]
\hline
\end{tabular}
\label{table:correlSocioEcoCT} 
\end{table}

To verify, we look at the Spearman correlation between deprivation, inequality, and segregation for the centroid city definition of each class, as indicated also in figure \ref{fig:zonesDefs}. We find only one correlation with a $R^2$ over 5\% (in bold in table \ref{table:correlSocioEcoCT}): the negative correlation between inequality and segregation in the 'Dispersed' class (i.e. the class which is the furthest away from plausible definitions of cities). This result enables us to consider deprivation, inequality and segregation as three independent dimensions to characterise cities.

\section{Results}

\subsection{Distributions of correlation coefficients for all 4914 city definitions}\label{subsubsec:distr}

\begin{figure}[h!]
\centering
\caption{Distributions of the Spearman correlation coefficient for the relation between EDI and a selection of mobile phone indicators calculated for all 4914 city definitions. The histogram is colored green when correlation coefficients are positive and orange when negative.}
\label{fig:group_corr_EDI_selection}
\resizebox{0.8\textwidth}{!}{ 
\includegraphics{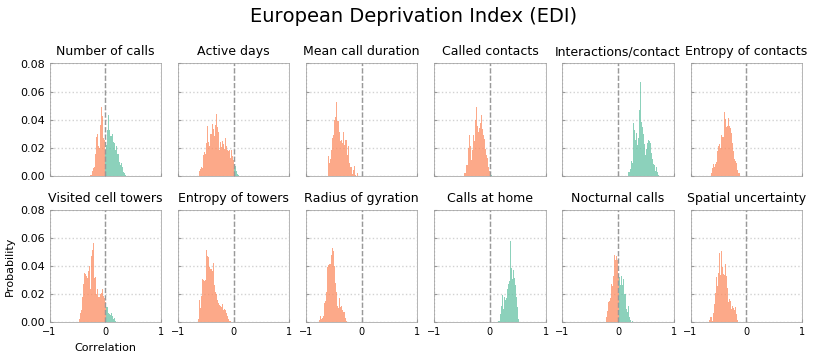}
}
\end{figure}

Visualising the distributions of obtained Spearman correlation coefficients for all 4914 city definitions allows us to (partly) assess the sensitivity of relations to city definitions. For the relation between EDI and several mobile phone indicators, for example, figure \ref{fig:group_corr_EDI_selection} shows the distribution of Spearman correlation coefficients over all city definitions. The figure suggests that for the relation between EDI and some mobile phone indicators (interactions per contact, percentage of calls at home, mean call duration), altering the city definition does not affect the direction of the correlation computed, although differences in significance occur for different city definitions. Remarkable is that for some mobile phone indicators, the relation with EDI can change direction depending on the city definition. One example is the relation between EDI and the number of calls in \ref{fig:group_corr_EDI_selection}. For this relation, a part of the city definitions results in positive correlation coefficients (meaning that the residents of the poorest cities, according to these definitions, have called more) but a another part of the city definitions results in negative correlation coefficients (meaning that the residents of the poorest cities have called less). The relation between EDI and mobile phone indicators is thus influenced by city definition, leading to differences in significance or even to changes in correlation directions between city definitions. 

\begin{figure}[h!]
\centering
\caption{Distributions of the Spearman correlation coefficient for the relation between the Gini index and a selection of mobile phone indicators calculated for all 4914 city definitions. The histogram is coloured green when correlation coefficients are positive and orange when negative.}
\label{fig:group_corr_GINI}
\resizebox{0.8\textwidth}{!}{ 
\includegraphics{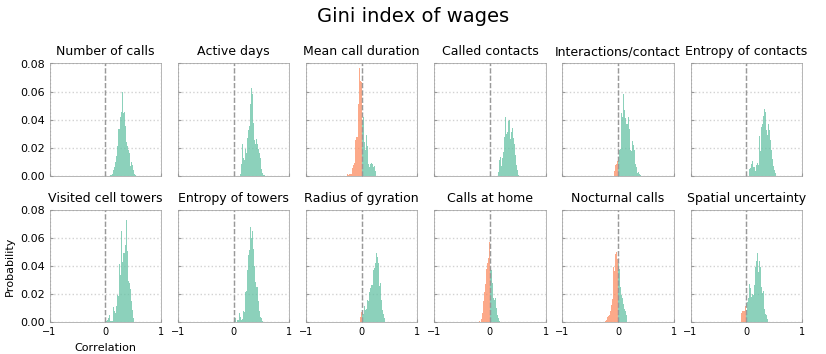}
} 
\end{figure}

Regarding the correlation of mobile phone indicators with the Gini index of wages inequality (figure \ref{fig:group_corr_GINI}), we find more robust relationships across city definitions, with the human activity sensed by mobile phone generally positively correlated with inequality. For example, the number of calls, their diversity (entropy of contacts) and the mobility range and diversity tend to increase in cities with a larger level of inequality (generally larger cities, cf. figure \ref{fig:au_maps_socioeco}). Only call-specific indicators are rather uncorrelated with inequality.

\begin{figure}[h!]
\centering
\caption{Distributions of the Spearman correlation coefficient for the relation between the Segregation index and a selection of mobile phone indicators calculated for all 4914 city definitions. The histogram is colored green when correlation coefficients are positive and orange when negative.}
\label{fig:group_corr_SEG}
\resizebox{0.8\textwidth}{!}{ 
\includegraphics{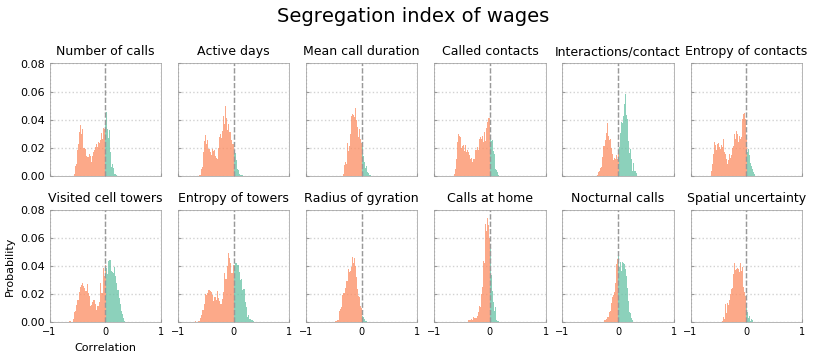}
} 
\end{figure}

Regarding the correlation of mobile phone indicators with the segregation index of wages (figure \ref{fig:group_corr_SEG}), we find that most human activity sensed by mobile phone tend to vary negatively with segregation. For example, the number of calls, their diversity (entropy of contacts) and the mobility range and diversity tend to decrease in cities with a larger level of segregation. However, these trends are more mixed and the choice of urban definition affects the sign of the coefficient obtained.

\subsection{Distributions of correlation coefficients by definition cluster}\label{subsubsec:distr_per_cluster}

\begin{figure}[h!]
\centering
\caption{Distributions of the Spearman correlation coefficient for the relation between the Entropy of contacts and the different socioeconomic indicators. Correlation coefficients are calculated for all 4914 city definitions but the histograms group results by the different classes of city definitions as defined in figure \ref{fig:zonesDefs}. The bars in the histograms are coloured green when correlation coefficients are positive and orange when negative. The colours outlining the histograms accord to the different classes as defined in figure \ref{fig:zonesDefs}.}
\label{fig:group_corr_entropy_of_contacts}
\resizebox{1\textwidth}{!}{ 
\includegraphics{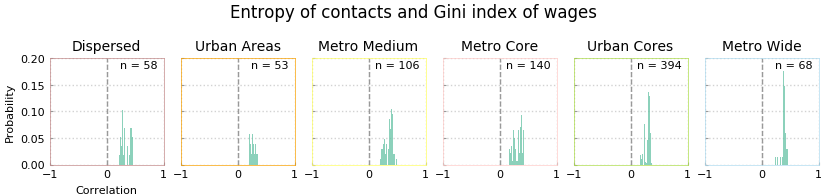}
} 
\resizebox{1\textwidth}{!}{ 
\includegraphics{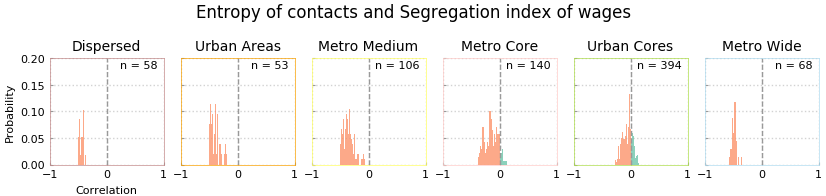}
} 
\resizebox{1\textwidth}{!}{ 
\includegraphics{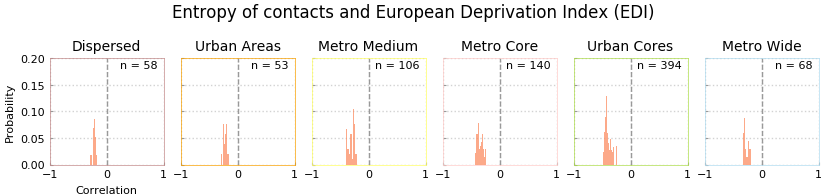}
} 
\end{figure}

Another outstanding question is whether the correlations between indicators are sensitive to the six classes of city definitions defined earlier in figure \ref{fig:zonesDefs}. To investigate this, figure \ref{fig:group_corr_entropy_of_contacts} shows the correlation coefficient, in this case for the relation between Entropy of Contacts and the socioeconomic indicators of cities, for all city definitions belonging to one of the six classes. In the case of the relation between Entropy of contacts and the Gini index, we find that correlations are similar for all six classes of city definitions. This is also the case for the relation between Entropy of contact, although here, the MetroCore and UrbanCore classes seem to diverge from the other classes as their histogram shows both positive and negative relations found, both with rather limited significance. For the relation between Entropy of contacts and EDI results are rather similar across all six classes, but again with limited significance.

\clearpage
\subsubsection{Heatmaps of correlations}\label{subsubsec:heatmaps}

The outstanding question therefore becomes which city definitions leads towards which correlations? 

\begin{figure}[htbp!]
\medskip
\centering
\includegraphics{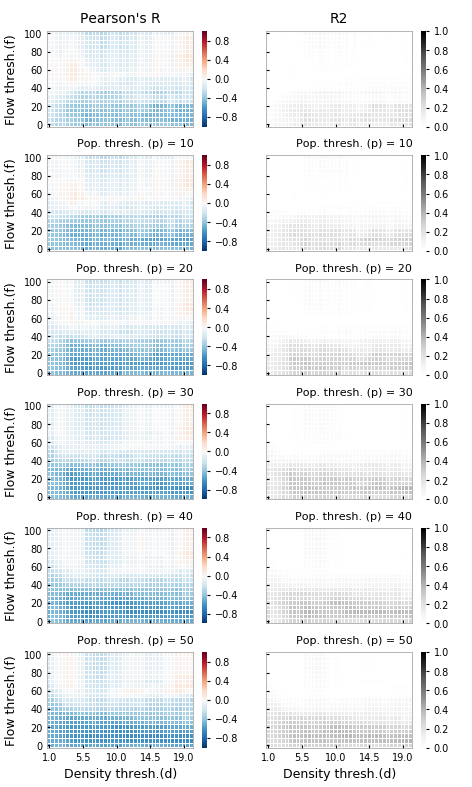}
\caption[Correlations between the entropy of contacts and the segregation index of wages in the city definition parameter-space]{Heatmap of the Pearsons's R (left) and $R^2$ values (right) for the relation between the entropy of contacts and the segregation index of wages for all city definitions represented in their according parameter-space. Each box represents one of the 4914 city definitions. Density thresholds ($d$) are for the city centers and in thousands inhabitants/hectare, flow thresholds (f) are in percentage of population commuting to the city center, and population thresholds (p) are in thousands inhabitants in the wider city. As can be deduced, the top row plots have a population threshold ($p$) of 0.}
\label{fig:heatmap_corr_entrcontact_SEG}
\end{figure}

An answer to this question can be formulated by mapping the obtained correlations coefficients between two indicators to the parameter-space used for the city definition. The heatmap in figure~\ref{fig:heatmap_corr_entrcontact_SEG}, for example, does so for the relation between the entropy of contacts and the segregation index of wages. Overall, results show that the entropy of mobile phone contacts is inversely related to wage segregation but different combinations of thresholds influence the observed correlations, their strength, and their significance. In other words it seems that less diverse mobile calling occurs together with higher segregation measures, but such relations only gain significance when cities are defined by low flow thresholds (f), so for rather loose definitions of suburban regions.

\clearpage

\subsection{Multiple regression of mobile phone indices with socioeconomic indicators }
\label{subsubsec:multiple_regression}

In order to combine the explaining power of all three aspect of the socioeconomic organisation of urban clusters, we have regressed the value of each aggregated mobile indicators with the value of deprivation, inequality and segregation for each definition of cities. In this section, we report the results for the 6 centres of classes defined previously (with a population cutoff of 10,000 residents) only for significance levels above 0.05.\\

\begin{figure}[h!]
\centering
\caption{Significant coefficients in a multiple regression of mobile indicators by cluster medoid. NB: In this figure, the number of observation N refers to the number of clusters within the representative medoid definition. It is a number of cities which are included in the regression for a given definition.}
\label{fig:multireg}
\resizebox{1\textwidth}{!}{ 
\includegraphics{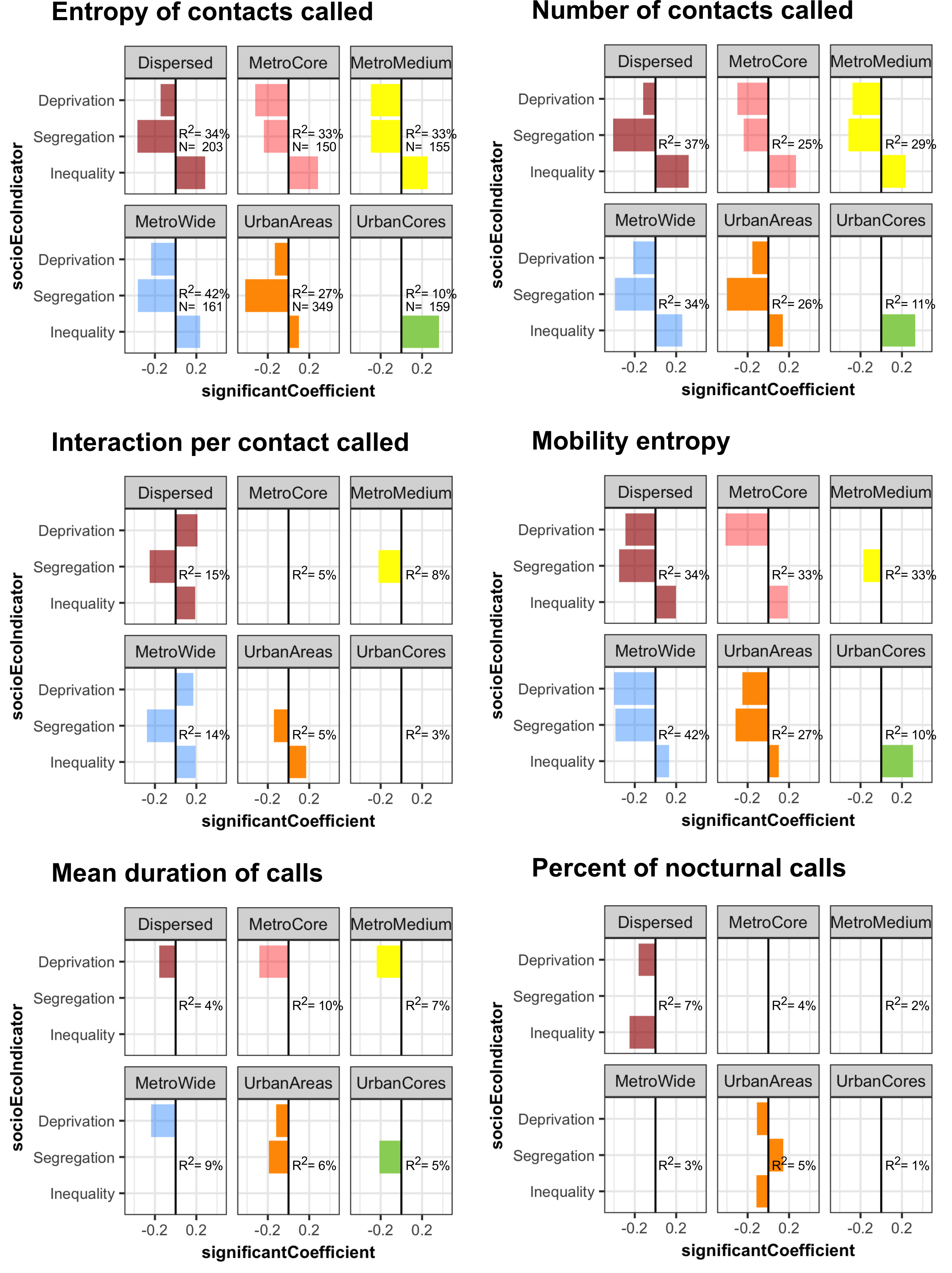}
}
\end{figure}

Most social behaviours as sensed by mobile phone activity and aggregated at the city level seem to be influenced by (or at least correlate to) the three socioeconomic dimensions cities (figure \ref{fig:multireg}). For the number and diversity of contacts called for example, we see that, regardless of the city definition used, socioeconomic indicators work in the same direction: more deprivation and more segregation lessen significantly and on average the number of contacts called and their diversity, whereas higher inequality in the city has an opposite result. There is sociological evidence relating to the reduction of social networks with individual and neighbourhood deprivation \cite{cattell2001poor} which could explain the statistical relations observed in this case. Combined with the observation that in some cases (clusters corresponding to very large urban perimeters rather than dense urban cores), deprivation correlates positively with the intensity of contacts (interaction per contact called: middle left graphs of figure \ref{fig:multireg}), this could match the usual observation that poorer actors have networks composed of more strong ties (more intense relations) and less weak ties (less diverse relations) than richer actors \citep{granovetter1983strength}. Higher deprivation and a stronger spatial concentration of wages could thus reduce the size and diversity of the social network with which an average individual interact virtually through mobile contacts. The important thing here is that, despite the mobile phone data dating from 2007, it might not just be the effect of the repelling cost of calls, because it is probably the cause behind the negative coefficient of the deprivation index on the mean duration of calls but in this latter case, segregation plays no role for most city definitions. \\

Finally, we find it intriguing that urban level of wage inequality would foster the number and diversity of contacts. This might be an effect of higher professional interdependency between the richest and the poorest in more unequal cities \cite{eeckhout2014spatial}... or simply an indirect effect of city size (which correlates positively with wage inequality, cf. figure \ref{fig:au_maps_socioeco}). In this case, the larger pool of potential contacts would increase the average actual diversity of contacts of individuals.\\

It is interesting to note that the intensity of the coefficients of the multiple regression vary slightly between the different clusters of city definitions but the overall picture is the same, except in the urban core clusters (the one composed of very dense city cores with little or no commuting periphery). Using these types of definition, the only significant variables at play is the level of inequality, which covariates positively with the entropy and number of contacts calls. The absence of effect of deprivation and segregation in French dense city cores could indicate that the centrality and density of the residence has a positive effect on the size and diversity of the social network which is reflected in the phone behaviour observed. More generally, the predictive power of regressions ($R^2$) is always the weakest for urban cores, whereas it peaks for definitions of clusters as metropolitan areas and dispersed areas. This suggests that activity behaviours of the inner parts of metropolises, as sensed by mobile phones throughout the day, cannot be well described by any of their static socioeconomic properties.\\

As for the physical mobility behaviours sensed by mobile phone activity and aggregated at the city level, we see the exact same pattern as for the number and diversity of contacts. Therefore, the network of physical encounters seems to be influenced by the same variables as the social network of contacts, which is an interesting result.\\

Finally, we chose the mean duration of calls and the percentage of nocturnal calls to show that not all mobile phone behaviours covariate with the socioeconomic structure of cities (cf. the values of $R^2<10\%$ in bottom graphs of figure \ref{fig:multireg}, whereas most other multiple regressions reach between 10 and 40\%). The percentage of nocturnal calls for example is orthogonal to all three indicators for most city definitions, whereas the mean duration of calls seems to be only significantly impacted by the mean deprivation level of cities.

\clearpage
\section{Discussion \& Conclusion}

In this paper, we have tried to take advantage of small and big data to bridge a gap between what is known about 'night-time' residential socioeconomic characteristics of urban areas in France, their 'day-time' production of inequality and segregation through wages, and the average social and spatial networks of citizens sensed with passive mobile phone data. We did so using French municipalities as a common unit of aggregation. We then studied the effect of city definition on the distribution and correlations of indicators obtained, building on previous work on the impact of city delineation on urban scaling results. Indeed, even though there is evidence of geographic concentration of inequalities with city size \citep{sarkar2018urban}, it was showed that this statistical relation varied with the city delineation chosen \citep{Cottineau2018Defining} and did not hold for other aspects of inequality such as spatial segregation. In the present paper, we add to this evidence by showing that the fixed socioeconomic characteristics of urban clusters also do not relate monotonically with sociospatial activities sensed by mobile phone data. For example, segregation can be positively or negatively correlated with the average diversity of places visited and the number of interactions per contact, depending on the urban delineation chosen among about 5000 possible ones based on variations of density, commuting and total population criteria. Moreover, we have found that the combination of our three socioeconomic indicators was more or less predictive of social and spatial activity levels. For example, high numbers and diversity of contacts are "explained" for a third of their variation by low levels of deprivation and segregation, and a higher inequality (economic diversity) in cities, especially when they are considered in their functional delineation of metropolitan areas (i.e. including commuters). However, social activities such as the duration of calls and their nocturnal aspect are left unexplained by the socioeconomic organisation of cities. \\

This study has enabled us to further assess the quality of mobile phone data for social science, to examine its relationship with traditional small data such as census and administrative data, as well as to look at their geographical variability. However, it is prone to some biases, such as the aggregation of data at the cell tower and resulting Modifiable Area Unit Problem (MAUP) effects (plus the fact that dense areas are better described, with more antennas, than rural areas). Furthermore, the data we have used are rather old (2007) and hard to update at this scale (six months worth of sensing) given the new laws regarding individual data privacy. Finally, mobile phone behaviours have changed with the widespread smartphones and mobile data usage, which makes this study hard to replicate nowadays.\\

However, our point is to emphasize that observed correlations of geographical data need to account for delineation sensitivity or to justify why a specific spatial delineation is preferred over others. In the absence of such justification, the exploration of delineations helps highlighting robust correlations (which work in all configurations), systematic variations (which respond to some characteristics of the urban space taken into account) and random variations (which prove the spuriousness of correlations reported on a single spatial delineations).

\bibliographystyle{apalike}
\bibliography{Paper_mobind_socecon.bib}

\begin{thebibliography}{-------}
\providecommand{\natexlab}[1]{#1}

\bibitem[Gonzalez \em{et~al.}(2008)Gonzalez, Hidalgo, and
  Barabasi]{gonzalez2008understanding}
Gonzalez, M.C.; Hidalgo, C.A.; Barabasi, A.L.
\newblock {Understanding individual human mobility patterns}.
\newblock {\em nature} {\bf 2008}, {\em 453},~779.

\bibitem[Zhong \em{et~al.}(2016)Zhong, Batty, Manley, Wang, Wang, Chen, and
  Schmitt]{zhong2016variability}
Zhong, C.; Batty, M.; Manley, E.; Wang, J.; Wang, Z.; Chen, F.; Schmitt, G.
\newblock {Variability in regularity: Mining temporal mobility patterns in
  London, Singapore and Beijing using smart-card data}.
\newblock {\em PloS one} {\bf 2016}, {\em 11},~e0149222.

\bibitem[Lenormand \em{et~al.}(2015)Lenormand, Louail, Cant{\'{u}}-Ros,
  Picornell, Herranz, Arias, Barthelemy, {San Miguel}, and
  Ramasco]{lenormand2015influence}
Lenormand, M.; Louail, T.; Cant{\'{u}}-Ros, O.G.; Picornell, M.; Herranz, R.;
  Arias, J.M.; Barthelemy, M.; {San Miguel}, M.; Ramasco, J.J.
\newblock {Influence of sociodemographic characteristics on human mobility}.
\newblock {\em Scientific reports} {\bf 2015}, {\em 5},~10075.

\bibitem[{De Montjoye} \em{et~al.}(2015){De Montjoye}, Radaelli, Singh, and
  Others]{de2015unique}
{De Montjoye}, Y.A.; Radaelli, L.; Singh, V.K.; Others.
\newblock {Unique in the shopping mall: On the reidentifiability of credit card
  metadata}.
\newblock {\em Science} {\bf 2015}, {\em 347},~536--539.

\bibitem[Alessandretti \em{et~al.}(2018)Alessandretti, Sapiezynski, Sekara,
  Lehmann, and Baronchelli]{alessandretti2018evidence}
Alessandretti, L.; Sapiezynski, P.; Sekara, V.; Lehmann, S.; Baronchelli, A.
\newblock {Evidence for a conserved quantity in human mobility}.
\newblock {\em Nature Human Behaviour} {\bf 2018}, {\em 2},~485--491.
\newblock
  doi:{\changeurlcolor{black}\href{https://doi.org/10.1038/s41562-018-0364-x}{\detokenize{10.1038/s41562-018-0364-x}}}.

\bibitem[Pappalardo \em{et~al.}(2015)Pappalardo, Simini, Rinzivillo, Pedreschi,
  Giannotti, and Barab{\'{a}}si]{pappalardo2015returners}
Pappalardo, L.; Simini, F.; Rinzivillo, S.; Pedreschi, D.; Giannotti, F.;
  Barab{\'{a}}si, A.L.
\newblock {Returners and explorers dichotomy in human mobility}.
\newblock {\em Nature Communications} {\bf 2015}, {\em 6},~8166.
\newblock
  doi:{\changeurlcolor{black}\href{https://doi.org/10.1038/ncomms9166}{\detokenize{10.1038/ncomms9166}}}.

\bibitem[Batran \em{et~al.}(2018)Batran, Mejia, Kanasugi, Sekimoto, and
  Shibasaki]{batran2018inferencing}
Batran, M.; Mejia, M.; Kanasugi, H.; Sekimoto, Y.; Shibasaki, R.
\newblock {Inferencing Human Spatiotemporal Mobility in Greater Maputo via
  Mobile Phone Big Data Mining}.
\newblock {\em ISPRS International Journal of Geo-Information} {\bf 2018}, {\em
  7},~259.

\bibitem[Lu \em{et~al.}(2017)Lu, Fang, Zhang, Shaw, Yin, Zhao, and
  Yang]{lu2017understanding}
Lu, S.; Fang, Z.; Zhang, X.; Shaw, S.L.; Yin, L.; Zhao, Z.; Yang, X.
\newblock {Understanding the representativeness of mobile phone location data
  in characterizing human mobility indicators}.
\newblock {\em ISPRS International Journal of Geo-Information} {\bf 2017}, {\em
  6},~7.

\bibitem[Vanhoof \em{et~al.}(2018)Vanhoof, Reis, Ploetz, and
  Smoreda]{Vanhoof2018HomeDet}
Vanhoof, M.; Reis, F.; Ploetz, T.; Smoreda, Z.
\newblock {Assessing the quality of home detection from mobile phone data for
  official statistics}.
\newblock {\em Journal of Official Statistics} {\bf 2018}, {\em In
  Press},~1--30,  \href{http://xxx.lanl.gov/abs/arXiv:1809.07567}{{\normalfont
  [arXiv:1809.07567]}}.

\bibitem[Longley \em{et~al.}(2015)Longley, Adnan, and
  Lansley]{longley2015geotemporal}
Longley, P.A.; Adnan, M.; Lansley, G.
\newblock {The geotemporal demographics of Twitter usage}.
\newblock {\em Environment and Planning A} {\bf 2015}, {\em 47},~465--484.

\bibitem[Arai \em{et~al.}(2016)Arai, Fan, Matekenya, and
  Shibasaki]{arai2016comparative}
Arai, A.; Fan, Z.; Matekenya, D.; Shibasaki, R.
\newblock {Comparative perspective of human behavior patterns to uncover
  ownership bias among mobile phone users}.
\newblock {\em ISPRS International Journal of Geo-Information} {\bf 2016}, {\em
  5},~85.

\bibitem[Schneider \em{et~al.}(2013)Schneider, Belik, Couronn{\'{e}}, Smoreda,
  and Gonz{\'{a}}lez]{schneider2013unravelling}
Schneider, C.M.; Belik, V.; Couronn{\'{e}}, T.; Smoreda, Z.; Gonz{\'{a}}lez,
  M.C.
\newblock {Unravelling daily human mobility motifs}.
\newblock {\em Journal of The Royal Society Interface} {\bf 2013}, {\em
  10},~20130246.

\bibitem[Dannamann \em{et~al.}(2018)Dannamann, Sotomayor-G{\'{o}}mez, and
  Samaniego]{dannamann2018time}
Dannamann, T.; Sotomayor-G{\'{o}}mez, B.; Samaniego, H.
\newblock {The time geography of segregation during working hours}.
\newblock {\em arXiv preprint arXiv:1802.00117} {\bf 2018}.

\bibitem[Xu \em{et~al.}(2018)Xu, Belyi, Bojic, and Ratti]{xu2018human}
Xu, Y.; Belyi, A.; Bojic, I.; Ratti, C.
\newblock {Human mobility and socioeconomic status: Analysis of Singapore and
  Boston}.
\newblock {\em Computers, Environment and Urban Systems} {\bf 2018}, {\em
  72},~51--67.
\newblock
  doi:{\changeurlcolor{black}\href{https://doi.org/10.1016/j.compenvurbsys.2018.04.001}{\detokenize{10.1016/j.compenvurbsys.2018.04.001}}}.

\bibitem[Vanhoof \em{et~al.}(2018{\natexlab{a}})Vanhoof, Schoors, {Van
  Rompaey}, Ploetz, and Smoreda]{Vanhoof2018MobEntr}
Vanhoof, M.; Schoors, W.; {Van Rompaey}, A.; Ploetz, T.; Smoreda, Z.
\newblock {Comparing Regional Patterns of Individual Movement Using Corrected
  Mobility Entropy}.
\newblock {\em Journal of Urban Technology} {\bf 2018}, {\em 25},~27--61.
\newblock
  doi:{\changeurlcolor{black}\href{https://doi.org/10.1080/10630732.2018.1450593}{\detokenize{10.1080/10630732.2018.1450593}}}.

\bibitem[Vanhoof \em{et~al.}(2018{\natexlab{b}})Vanhoof, Lee, and
  Smoreda]{Vanhoof2018Sensitivities}
Vanhoof, M.; Lee, C.; Smoreda, Z.
\newblock {Performance and sensitivities of home detection from mobile phone
  data} {\bf 2018}.
\newblock pp. 1--18,
  \href{http://xxx.lanl.gov/abs/arXiv:1809.09911}{{\normalfont
  [arXiv:1809.09911]}}.

\bibitem[Pappalardo \em{et~al.}(2016)Pappalardo, Vanhoof, Gabrielli, Smoreda,
  Pedreschi, and Giannotti]{Pappalardo2016}
Pappalardo, L.; Vanhoof, M.; Gabrielli, L.; Smoreda, Z.; Pedreschi, D.;
  Giannotti, F.
\newblock {An analytical framework to nowcast well-being using mobile phone
  data}.
\newblock {\em International Journal of Data Science and Analytics} {\bf 2016},
  {\em 2},~75--92.
\newblock
  doi:{\changeurlcolor{black}\href{https://doi.org/10.1007/s41060-016-0013-2}{\detokenize{10.1007/s41060-016-0013-2}}}.

\bibitem[Eagle \em{et~al.}(2010)Eagle, Macy, and Claxton]{Eagle2010}
Eagle, N.; Macy, M.; Claxton, R.
\newblock {Network diversity and economic development}.
\newblock {\em Science} {\bf 2010}, {\em 328},~1029--1031,
  \href{http://xxx.lanl.gov/abs/1011.0208}{{\normalfont [1011.0208]}}.
\newblock
  doi:{\changeurlcolor{black}\href{https://doi.org/10.1126/science.1186605}{\detokenize{10.1126/science.1186605}}}.

\bibitem[Decuyper \em{et~al.}(2014)Decuyper, Rutherford, Wadhwa, Bauer, Krings,
  Gutierrez, Blondel, and Luengo-Oroz]{Decuyper2014}
Decuyper, A.; Rutherford, A.; Wadhwa, A.; Bauer, J.M.; Krings, G.; Gutierrez,
  T.; Blondel, V.D.; Luengo-Oroz, M.A.
\newblock {Estimating food consumption and poverty indices with mobile phone
  data} {\bf 2014}.
\newblock  \href{http://xxx.lanl.gov/abs/1412.2595}{{\normalfont [1412.2595]}}.

\bibitem[Frias-martinez \em{et~al.}(2013)Frias-martinez, Soto, Virseda, and
  Frias-martinez]{Frias-martinez2013}
Frias-martinez, V.; Soto, V.; Virseda, J.; Frias-martinez, E.
\newblock {Can cell phone traces measure social development ?}
\newblock  Third Conference on the Analysis of Mobile Phone datasets, NetMob;
  {Vincent Blondel}.; {Adeline Decuyper}.; Pierre, D.; {Yves-Alexandre, De
  Montjoye Jameson}, T.; Vincent, T.; Dashun, W., Eds.; ,  2013; pp. 62--65.

\bibitem[Arcaute \em{et~al.}(2015)Arcaute, Hatna, Ferguson, Youn, Johansson,
  and Batty]{Arcaute2015}
Arcaute, E.; Hatna, E.; Ferguson, P.; Youn, H.; Johansson, A.; Batty, M.
\newblock {Constructing cities, deconstructing scaling laws.}
\newblock {\em Journal of the Royal Society, Interface} {\bf 2015}, {\em
  12},~20140745.
\newblock
  doi:{\changeurlcolor{black}\href{https://doi.org/10.1098/rsif.2014.0745}{\detokenize{10.1098/rsif.2014.0745}}}.

\bibitem[Veneri(2016)]{veneri2016city}
Veneri, P.
\newblock City size distribution across the OECD: Does the definition of cities
  matter?
\newblock {\em Computers, Environment and Urban Systems} {\bf 2016}, {\em
  59},~86--94.

\bibitem[Cottineau \em{et~al.}(2017)Cottineau, Hatna, Arcaute, and
  Batty]{cottineau2017diverse}
Cottineau, C.; Hatna, E.; Arcaute, E.; Batty, M.
\newblock {Diverse cities or the systematic paradox of Urban Scaling Laws}.
\newblock {\em Computers, Environment and Urban Systems} {\bf 2017}, {\em
  63},~80--94.

\bibitem[Pornet \em{et~al.}(2012)Pornet, Delpierre, Dejardin, Grosclaude,
  Launay, Guittet, Lang, and Launoy]{Pornet2012}
Pornet, C.; Delpierre, C.; Dejardin, O.; Grosclaude, P.; Launay, L.; Guittet,
  L.; Lang, T.; Launoy, G.
\newblock {Construction of an adaptable European transnational ecological
  deprivation index: The French version}.
\newblock {\em Journal of Epidemiology and Community Health} {\bf 2012}, {\em
  66},~982--989.
\newblock
  doi:{\changeurlcolor{black}\href{https://doi.org/10.1136/jech-2011-200311}{\detokenize{10.1136/jech-2011-200311}}}.

\bibitem[Cottineau \em{et~al.}(2018)Cottineau, Finance, Hatna, Arcaute, and
  Batty]{Cottineau2018Defining}
Cottineau, C.; Finance, O.; Hatna, E.; Arcaute, E.; Batty, M.
\newblock {Defining urban clusters to detect agglomeration economies}.
\newblock {\em Environment and Planning B: Urban Analytics and City Science}
  {\bf 2018}.
\newblock
  doi:{\changeurlcolor{black}\href{https://doi.org/10.1177/2399808318755146}{\detokenize{10.1177/2399808318755146}}}.

\bibitem[Fuller(1979)]{fuller1979estimation}
Fuller, M.
\newblock {The estimation of Gini coefficients from grouped data: Upper and
  Lower Bounds}.
\newblock {\em Economics Letters} {\bf 1979}, {\em 3},~187--192.
\newblock
  doi:{\changeurlcolor{black}\href{https://doi.org/10.1016/0165-1765(79)90115-0}{\detokenize{10.1016/0165-1765(79)90115-0}}}.

\bibitem[Reardon(2009)]{reardon2009measures}
Reardon, S.F.
\newblock {Measures of ordinal segregation}. In {\em Occupational and
  Residential Segregation}, Researcg o ed.;  Fl{\"{u}}ckiger, Y.; Reardon,
  S.F.; Silber, J., Eds.; Emeral Group Publishing Limited,  2009; pp. 129--155.
\newblock
  doi:{\changeurlcolor{black}\href{https://doi.org/10.1108/S1049-2585(2009)0000017011}{\detokenize{10.1108/S1049-2585(2009)0000017011}}}.

\bibitem[Grauwin \em{et~al.}(2017)Grauwin, Szell, Sobolevsky, H{\"{o}}vel,
  Simini, Vanhoof, Smoreda, Barab{\'{a}}si, and Ratti]{Grauwin2017}
Grauwin, S.; Szell, M.; Sobolevsky, S.; H{\"{o}}vel, P.; Simini, F.; Vanhoof,
  M.; Smoreda, Z.; Barab{\'{a}}si, A.L.; Ratti, C.
\newblock {Identifying and modeling the structural discontinuities of human
  interactions}.
\newblock {\em Scientific Reports} {\bf 2017}, {\em 7},~46677.
\newblock
  doi:{\changeurlcolor{black}\href{https://doi.org/10.1038/srep46677}{\detokenize{10.1038/srep46677}}}.

\bibitem[Vanhoof \em{et~al.}(2017{\natexlab{a}})Vanhoof, Hendrickx, Puussaar,
  Verstraeten, Ploetz, and Smoreda]{Vanhoof2017domestictourism}
Vanhoof, M.; Hendrickx, L.; Puussaar, A.; Verstraeten, G.; Ploetz, T.; Smoreda,
  Z.
\newblock {Exploring the use of mobile phones during domestic tourism trips}.
\newblock {\em Netcom} {\bf 2017}, {\em 31},~335--372.

\bibitem[Vanhoof \em{et~al.}(2017{\natexlab{b}})Vanhoof, Combes, and {De
  Bellefon}]{Vanhoof2017miningurbanareas}
Vanhoof, M.; Combes, S.; {De Bellefon}, M.P.
\newblock {Mining mobile phone data to detect urban areas}.
\newblock  Statistics and Data Science: New challenges, new generations, SIS
  2017; Petrucci, A.; Verde, R., Eds.; Firenze University Press: Firenze,
  2017; pp. 1005--1012.

\bibitem[{De Montjoye} \em{et~al.}(2016){De Montjoye}, Rocher, and
  Pentland]{DeMontjoye2016}
{De Montjoye}, Y.A.; Rocher, L.; Pentland, A.S.
\newblock {Bandicoot: a python toolbox for mobile phone metadata}.
\newblock {\em The Journal of Machine Learning Research} {\bf 2016}, {\em
  17},~6100--6104.

\bibitem[Cottineau(2017)]{cottineau2017metazipf}
Cottineau, C.
\newblock {MetaZipf. A dynamic meta-analysis of city size distributions}.
\newblock {\em PloS one} {\bf 2017}, {\em 12},~e0183919.

\bibitem[Gower(1971)]{gower1971general}
Gower, J.C.
\newblock A general coefficient of similarity and some of its properties.
\newblock {\em Biometrics} {\bf 1971}, pp. 857--871.

\bibitem[Kaufman and Rousseeuw(1987)]{kaufman1987clustering}
Kaufman, L.; Rousseeuw, P.
\newblock {\em Clustering by means of medoids}; North-Holland,  1987.

\bibitem[Cattell(2001)]{cattell2001poor}
Cattell, V.
\newblock {Poor people, poor places, and poor health: the mediating role of
  social networks and social capital}.
\newblock {\em Social science {\&} medicine} {\bf 2001}, {\em 52},~1501--1516.

\bibitem[Granovetter(1983)]{granovetter1983strength}
Granovetter, M.
\newblock The strength of weak ties: A network theory revisited.
\newblock {\em Sociological theory} {\bf 1983}, pp. 201--233.

\bibitem[Eeckhout \em{et~al.}(2014)Eeckhout, Pinheiro, and
  Schmidheiny]{eeckhout2014spatial}
Eeckhout, J.; Pinheiro, R.; Schmidheiny, K.
\newblock {Spatial sorting}.
\newblock {\em Journal of Political Economy} {\bf 2014}, {\em 122},~554--620.

\bibitem[Sarkar(2018)]{sarkar2018urban}
Sarkar, S.
\newblock {Urban scaling and the geographic concentration of inequalities by
  city size}.
\newblock {\em Environment and Planning B: Urban Analytics and City Science}
  {\bf 2018}, p. 2399808318766070.

\end{thebibliography}

\section*{Acknowledgements}
We are grateful for our funding institutions at the time of this project (UCL, CNRS, University of Newcastle and Orange Labs). We also want to thank Elsa Arcaute for inspiring discussions at the start of this collaboration, as well as Clement Lee and Thomas Louail for their valuable comments on the manuscript.

\end{document}